\documentstyle[aps,prl,twocolumn,psfig,floats]{revtex}     

\begin{document}
\preprint{\vbox{\hbox{{\tt hep-ph/0111264}\\ November 2001}}}
\draft
\wideabs{
\title{Shadows of the Planck Scale:\\  The Changing Face of Compactification Geometry}
\author{Keith R. Dienes \,and\, Arash Mafi}
\address{Department of Physics, University of Arizona, Tucson, AZ  85721 USA}
\address{E-mail addresses:  ~{\tt dienes,mafi@physics.arizona.edu}}
\date{November 28, 2001}
\maketitle
\begin{abstract}
     By studying the effects of the shape moduli associated with 
     toroidal compactifications, we demonstrate that 
     Planck-sized extra dimensions can cast
     significant ``shadows'' over low-energy physics.
     These shadows can greatly distort our perceptions of 
     the compactification geometry associated with large extra dimensions,
     and place a fundamental limit on our ability to probe the 
     geometry of compactification simply by measuring Kaluza-Klein states.
          We also discuss the interpretation
          of compactification radii and hierarchies in the context 
          of geometries with non-trivial shape moduli.  
     One of the main results of this paper is that compactification geometry 
     is effectively renormalized as a function of energy scale, 
     with ``renormalization group equations'' describing the 
     ``flow'' of geometric parameters such as compactification radii and shape
     angles as functions of energy.
\end{abstract}
\bigskip
\bigskip
          }

\newcommand{\newc}{\newcommand}
\newc{\gsim}{\lower.7ex\hbox{$\;\stackrel{\textstyle>}{\sim}\;$}}
\newc{\lsim}{\lower.7ex\hbox{$\;\stackrel{\textstyle<}{\sim}\;$}}

\def\beq{\begin{equation}}
\def\eeq{\end{equation}}
\def\beqn{\begin{eqnarray}}
\def\eeqn{\end{eqnarray}}
\def\calM{{\cal M}}
\def\half{{\textstyle{1\over 2}}}
\def\ie{{\it i.e.}\/}
\def\eg{{\it e.g.}\/}


\def\inbar{\,\vrule height1.5ex width.4pt depth0pt}
\def\IR{\relax{\rm I\kern-.18em R}}
 \font\cmss=cmss10 \font\cmsss=cmss10 at 7pt
\def\IQ{\relax{\rm I\kern-.18em Q}}
\def\IZ{\relax\ifmmode\mathchoice
 {\hbox{\cmss Z\kern-.4em Z}}{\hbox{\cmss Z\kern-.4em Z}}
 {\lower.9pt\hbox{\cmsss Z\kern-.4em Z}}
 {\lower1.2pt\hbox{\cmsss Z\kern-.4em Z}}\else{\cmss Z\kern-.4em Z}\fi}

\input epsf


\section{Introduction}

In a recent paper~\cite{firstpaper},
it was shown that the shape moduli associated with
toroidal compactifications can have a number
of important effects on the corresponding Kaluza-Klein 
spectrum:  they induce level-crossing, they modify the mass gap,
and in certain cases they permit extra dimensions to grow infinitely large
without violating experimental constraints.
These results suggest that shape moduli have the potential to 
drastically change our na\"\i ve expectations based on studying
simple compactifications in which shape moduli are ignored or held fixed.

The purpose of this paper is to demonstrate several
further surprising consequences of shape moduli.
One of the crucial differences between volume moduli and shape moduli
is that volume moduli are {\it dimensionful}\/ and necessarily have
energy scales associated with them;  by contrast, shape moduli are
dimensionless.  
Thus, even when the radii and volume of certain extra dimensions
are taken to zero, 
the corresponding shape moduli do not vanish and can still play a significant
role in affecting our understanding of low-energy phenomenology.
Indeed, much like the smile of the Cheshire cat, 
the shape moduli can survive 
and ultimately distort our perceptions of purely low-energy physics ---
even when the extra dimensions to which they correspond
are no larger than the Planck length!

As we shall see, this observation implies that 
it is impossible to verify the ``true'' compactification
geometry experimentally --- indeed, the whole notion of a fixed
compactification geometry becomes experimentally meaningless.
By contrast, we shall show that
compactification geometry, much like other ``constants'' of nature,
is effectively {\it renormalized}\/ as a function of energy scale, 
with quantities such as compactification radii 
changing their apparent values as functions of the energy
with which the compactification manifold is probed.

\section{Correspondence Relations and Shadowing}

\begin{figure}[t]
\centerline{\epsfxsize 3.3 truein \epsfbox {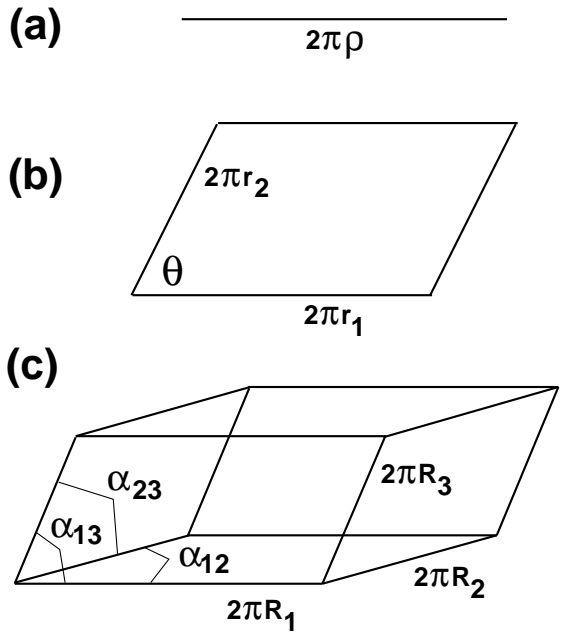}}
\caption{General one-, two-, and three-dimensional tori with arbitrary
     shape angles.}
\label{tori}
\end{figure}

In order to illustrate these ideas concretely,
let us consider compactifications on general one-, two-,
and three-dimensional tori.  These tori are
illustrated in Fig.~\ref{tori}, where opposite edges in
all diagrams are periodically identified.  
The one-dimensional torus, of course, is nothing but a circle
and has no corresponding shape moduli. 
By contrast, the two- and three-dimensional tori are described
not only by radii but also by the shift angles $\theta$
and $\alpha_{ij}$ which mix 
the periodicities associated with translations along the
corresponding directions.
In such cases, the shape moduli are the
shift angles as well as the ratios of the radii.
Note that tori with different shift angles are topologically
distinct (up to modular transformations which will be discussed
below).  However, despite the appearance of such shape moduli,
in all cases the compactification manifolds are {\it flat}\/.

Our goal is to study the extent to which various low-energy observers
can determine the shapes of these tori by studying their associated 
Kaluza-Klein spectra.  
Towards this end, let us assume that 
the ``true'' compactification geometry is given by the three-torus
shown in Fig.~\ref{tori}(c).  Furthermore, let us assume that 
there is a hierarchy of length scales such that $R_3\ll R_2\ll R_1$.
For example, $R_3$ might be near the Planck scale, while $R_1$ might
be at the inverse TeV scale and $R_2$ might be at some intermediate
scale. 
Of course, to a high-energy observer with access to energies 
$E_{\rm max}\gg {\cal O}(R_3^{-1})$, the Kaluza-Klein spectrum
will reveal the presence of all three dimensions of the torus.
Such an observer can then determine all three radii $R_i$ and 
shape angles $\alpha_{ij}$ through a detailed spectral analysis
of the Kaluza-Klein states.  
However, for an observer with 
access to only intermediate energies 
${\cal O}(R_2^{-1}) \ll E_{\rm max} \ll {\cal O}(R_3^{-1})$,
the third dimension will be inaccessible;  the compactification
manifold would then appear to be a two-torus, as illustrated
in Fig.~\ref{tori}(b).
Finally, for the low-energy observer with access to energies
${\cal O}(R_1^{-1}) \ll E_{\rm max} \ll {\cal O}(R_2^{-1})$,
only one dimension worth of Kaluza-Klein states will be accessible.
Such an observer would then conclude that the compactification
manifold is merely a circle, as illustrated in Fig.~\ref{tori}(a).

This change in the effective dimensionality of the compactification space
is obvious, and is not our focus in this paper.  However, given the 
hierarchy $R_3\ll R_2\ll R_1$,
it is natural to expect that the intermediate-energy observer
would experience a two-torus 
whose parameters $(r_1,r_2,\theta)$ are related to 
the underlying parameters $(R_i,\alpha_{ij})$ of the 
three-torus via 
\beq
          \cases{ \displaystyle {r_1 = R_1} &  \cr
                  \displaystyle{ r_2 = R_2 } &  \cr
                 \displaystyle{\theta = \alpha_{12}} & ~.\cr}  
\label{wrong1}
\eeq
After all, at energies much below $R_3^{-1}$, we expect all remnants
of the third dimension to vanish, so that the two-torus experienced
by the intermediate-energy observer is merely the ``base'' of the original
three-torus in Fig.~\ref{tori}(c).
Likewise, it is natural to expect that the lowest-energy observer 
would perceive a circle with radius $\rho=r_1$, which by Eq.~(\ref{wrong1})
implies
\beq
      \rho ~=~  r_1 ~=~ R_1  ~.
\label{wrong2}
\eeq
Once again, this would be the na\"\i ve 
expectation, given that we have only sufficient energy 
to probe the largest dimension of
the original three-torus.

The main point of this paper is to demonstrate that 
the correspondence relations in Eqs.~(\ref{wrong1}) and (\ref{wrong2}) 
are incorrect, even in the presence of a large hierarchy $R_3\ll R_2\ll R_1$, 
and must be replaced by relations which are far more non-trivial.  
Indeed, as we shall see,
the relations in Eqs.~(\ref{wrong1}) and (\ref{wrong2}) 
hold only when the shape moduli are ignored
(\ie, when all shape angles are taken to be $\pi/2$).
In the presence of non-rectangular shape angles, by contrast,
we shall see that these relations completely fail to describe the
process by which small extra dimensions can be ``integrated
out'' when passing to larger and larger length scales.
We stress that this failure occurs {\it no matter how small}\/ the
smallest radii become.  Thus, this failure can have
dramatic phenomenological consequences at low energies.
  
It is straightforward to determine the correct correspondence relations
by comparing the Kaluza-Klein spectra in each case.
In the case of the circle in Fig.~\ref{tori}(a), the Kaluza-Klein 
spectrum is given by 
\beq
            \calM^2 ~=~ {n_1^2 \over \rho^2}~,
\label{KKone}
\eeq
where $n_1\in\IZ$.
By contrast,
for the general two-torus shown in Fig.~\ref{tori}(b), the 
Kaluza-Klein spectrum is instead given by
\beq
        \calM^2 ~=~  {1\over \sin^2\theta} 
         \left\lbrack
                  \sum_{i=1}^2 {n_i^2 \over r_i^2}
             -   \sum_{i\not= j}  {n_i n_j \over r_i r_j} \cos\theta
          \right\rbrack~
\label{KKtwo}
\eeq
where $n_i\in \IZ$.
An explicit derivation of this result can be found, \eg, in Ref.~\cite{firstpaper};  
note that the periodicities of the two-torus allow us to restrict our
attention to the range $0\leq \theta\leq \pi/2$ without
loss of generality.  
Finally, for the general three-torus shown in Fig.~\ref{tori}(c), the 
Kaluza-Klein spectrum is given by~\cite{nextpaper}
\beq
        \calM^2 ~=~ {1\over K}\,
          \left\lbrack
           \sum_{i=1}^3 {n_i^2 \over R_i^2} s_{jk}^2 
           - \sum_{i\not= j}   {n_i n_j \over R_i R_j} 
             \left(c_{ij} - c_{ik} c_{jk} \right)
                \right\rbrack ~
\label{KKthree}
\eeq
where $k\not= i,j$, where 
$c_{ij}\equiv \cos\alpha_{ij}$
and 
$s_{ij}\equiv \sin\alpha_{ij}$,
and where $K$ (the dimensionless squared volume of the parallelepiped
in Fig.~\ref{tori}(c))
is given by
\beq
         K \equiv 1-  \sum_{i<j} c_{ij}^2 + 2 \prod_{i< j} c_{ij}
           =   
            \sum_{i<j} s_{ij}^2 + 2 \left(\prod_{i< j} c_{ij} -1 \right)~.
\eeq
Note that such a torus is physical 
only if $\alpha_{ij}+ \alpha_{jk} > \alpha_{ik}$ for all combinations of unequal $(i,j,k)$; 
this bound is saturated in the degenerate limit when one of the torus
periodicities lies in the plane of the other two.

Given these results, we can now determine the appropriate correspondence relations.
In deriving these relations, we shall assume a hierarchy $R_3\ll R_2\ll R_1$
so that we can successively integrate out small extra dimensions
when passing to larger length scales.
Our procedure will be to disregard all Kaluza-Klein
states whose masses exceed the appropriate reference energy 
(either high, intermediate, or low) and are therefore inaccessible to 
the corresponding observer.

The observer at highest energy clearly sees three dimensions worth
of Kaluza-Klein states, and deduces the ``true'' geometry of the
compactified space by comparing the measured Kaluza-Klein masses 
with Eq.~(\ref{KKthree}).
However, the observer at intermediate energy cannot perceive excitations
in the $R_3$ direction, since functionally $R_3\to 0$ for this
observer.  His attention is therefore restricted
to states with $n_3=0$, and he attempts a spectral analysis 
of the remaining states\footnote{
    This restriction to states with $n_3=0$ is applicable unless there exist
   special cancellations in the Kaluza-Klein mass formula.  
     We will discuss such special cases below, but they do
            not affect our results.}
via comparison with Eq.~(\ref{KKtwo}).
This leads to the identifications
\beqn
       {1\over \sin^2\theta} {1\over r_i^2} &=& {s_{jk}^2\over K R_i^2}~~~~~ (i=1,2)\nonumber\\
       {\cos\theta\over \sin^2\theta} {1\over r_1 r_2} &=&
           { c_{12} - c_{13}c_{23} \over K R_1 R_2}~. 
\eeqn
This observer therefore deduces that the compactified space
is a two-torus parametrized by $(r_1,r_2,\theta)$ 
given by
\beqn
   \cases{ \displaystyle{r_i}  &=  $\displaystyle{s_{i3}  R_i}$   \cr
           \displaystyle{\cos\theta}  & =  
             $\displaystyle{ (c_{12} - c_{13}c_{23}) / s_{13} s_{23} }$ ~.\cr} 
\label{right1}
\eeqn
Note that both the radii $r_i$ and the shape angle $\theta$ are affected, leading
to apparent values for $(r_1,r_2,\theta)$ which are not present in
the original three-torus.

The lowest-energy observer, by contrast, misses the $n_2$ excitations
as well.
Upon comparing with Eq.~(\ref{KKone}),
he therefore concludes that the compactified space is a circle 
of radius
\beq
        \rho ~=~ (\sin\theta) r_1 ~=~ {\sqrt{K} \over  s_{23}} \,R_1~. 
\label{right2}
\eeq
Once again, this radius does not correspond to any periodicity 
in the original three-torus.

Mathematically, these results 
reflect the geometric ``shadows''  
that successive smaller extra dimensions cast onto 
the larger extra dimensions when they are integrated out.
As such, they indicate that 
the low-energy observer can see only those ``projections''
of the compactification space which are perpendicular to 
the extinguished dimensions.
But given the assumed large hierarchy of length scales,
the physical implications of this shadowing effect are 
rather striking.
A small extra spacetime dimension --- even one no larger
than the Planck length! --- is able to cast a huge shadow
over all other length scales and their associated dimensions,
completely distorting our low-energy perception
and interpretation of the compactification geometry.
Indeed, the physics which we would normally associate with the
Planck scale (such as the angles that parametrize the shape
of the Planck-sized extra dimensions relative to the 
larger extra dimensions) fail to decouple
at low energies.

Let us consider an extreme example to illustrate this point.
If $\sqrt{K}/s_{23}\ll 1$ 
in Eq.~(\ref{right2}), then  $\rho$ can appear to be very
small at low energies even though $R_1$ itself might be huge.
Thus, the original three-torus would have a huge dimension $R_1$,
yet this dimension would be completely invisible at low energies because
of the ``shadow'' cast by the additional dimensions associated 
with $R_2$ and $R_3$.  This distortion occurs even though
these additional dimensions might be at the Planck scale!
While this is similar to the invisibility 
mechanism discussed in Ref.~\cite{firstpaper}, 
our point here is that the shadowing phenomenon is completely
general and holds even when $R_2$ and $R_3$ are vanishingly small. 
In other words, our normal expectations concerning decoupling 
do not apply when non-trivial shape moduli are involved.  
Other examples and situations will be discussed below.

Of course, no observer at any energy scale can use this shadowing
phenomenon in order to deduce the existence of an extra spacetime 
dimension beyond his own energy scale.
Nevertheless, the observer's {\it interpretation}\/ of 
that portion of the compactification geometry accessible to him
is completely distorted, leading him to 
deduce geometric radii and shape angles that have no basis 
in reality.  
Since the existence of an even
smaller extra dimension beyond those already perceived
can never be ruled out,
this shadowing effect implies that {\it one can 
never know the ``true'' compactification geometry}\/.
Even when light Kaluza-Klein states are detected and successful
fits to the Kaluza-Klein mass formulae are obtained via 
spectral analyses,
the presence of further additional dimensions with appropriate shape
moduli can always reveal the previous successes to have been
illusory.

We are not claiming that no ``true'' compactification
geometry can ever exist.  Indeed, if one takes the predictions of string
theory seriously, then there is ultimately a true, maximum number
of compactified dimensions, with associated radii and 
shape moduli.  However, as an {\it experimental}\/ question,
one can never be satisfied concerning the true number of extra
dimensions.  Thus, our result implies that one can correspondingly never 
be certain of the nature of whatever compactification geometry is ultimately
discovered.  In this sense, the concept of a ``true'' compactification
geometry does not exist.

\section{Modular transformations and compactification radii}

It is important to stress 
that the effects embodied in the
relations (\ref{right1}) and (\ref{right2})
cannot generally be undone either by changes of coordinate basis
or by modular transformations of the higher- or lower-dimensional tori.
Since the case of modular transformations is particularly important,
it merits some discussion.

Because of the modular symmetries of the torus, it is possible
to describe the topology of a given torus using a multitude 
of different values for the compactification radii and shape angles;  
only the corresponding Kaluza-Klein spectrum is physical and invariant
under modular transformations.  
Thus, modular transformations are
analogous to gauge transformations, providing redundant descriptions
of the same physics.  

Given this, the question then emerges as to whether the
shadowing effect can be undone via such modular transformations.  
Might there exist
an alternative, modular-equivalent description of the 
compactification radii and twist angles of either the 
original torus or the effective low-energy torus   
(or both)
such that the relations in Eqs.~(\ref{wrong1}) 
and (\ref{wrong2}) can be restored?  
It is relatively straightforward to show~\cite{nextpaper}
that the answer to this question is `no', and we shall
see an explicit example of this below.  
Thus, the effects of shadowing cannot be eliminated 
by exploiting modular symmetries;  they persist
no matter which modular-equivalent descriptions 
are used to describe the higher- or lower-dimensional tori.
In other words, they are {\it not}\/ ``pure-gauge''.

On the other hand, the existence of modular symmetries 
indicates that we must more carefully refine our concept
of ``compactification radii'' when discussing 
compactification manifolds with non-trivial shape moduli.  
In the case of toroidal compactifications, only the 
one-dimensional periodicity radius $\rho$ has an absolute meaning (since
circles have no associated modular symmetries);  the periodicity
radii of all higher-dimensional tori are not invariant
under modular transformations, and can be adjusted.
For example, a two-torus with parameters
$(r_1,r_2,\theta)$ is topologically the same 
(and thus has the same Kaluza-Klein spectrum) as the infinite
set of two-tori
with parameters $(r'_1,r'_2,\theta')$ given by
\beqn
    r_1' &=& r_2 \sqrt{ (c \cos\theta + dr_1/r_2)^2 + c^2 \sin^2\theta}\nonumber\\
    r_2' &=& r_2 \sqrt{ (a \cos\theta + br_1/r_2)^2 + a^2 \sin^2\theta}\nonumber\\
   \sin\theta' &=&   \sin\theta  \,(r_1 r_2 / r_1' r_2') 
\label{modtrans}
\eeqn
for all $a,b,c,d\in\IZ$ with $ad-bc=1$.

Given these equations, it is easy to verify that
no solution for $(a,b,c,d)$ can transform Eq.~(\ref{right2}) 
into the unshadowed result $\rho = r_1'$ or $\rho=r_2'$.
Specifically, there exists no solution for $(a,b,c,d)$ in Eq.~(\ref{modtrans})
yielding $r_1'= r \sin\theta$ or $r_2'=r\sin\theta$  
for general $r_1$, $r_2$, and $\theta$.
This explicitly demonstrates that shadowing
is a physical effect rather than a modular (gauge) artifact.

However, note that the ratio $r_2/r_1$ is not modular invariant;
this ratio can be adjusted 
even though the corresponding Kaluza-Klein spectrum is unaltered.
For example, if $r_2/r_1=1$ with $\theta=\pi/2$, we can set
$r_2'/r_1' = \sqrt{1+b^2}$ for arbitrarily large $b\in\IZ$
simply by taking $a=d=1$ and $c=0$.  Such a hierarchy is clearly
unphysical. 
How then can we properly define the notion of ``hierarchy''
that we have exploited in this paper?

\begin{figure}[h]
\centerline{\epsfxsize 3.5 truein \epsfbox {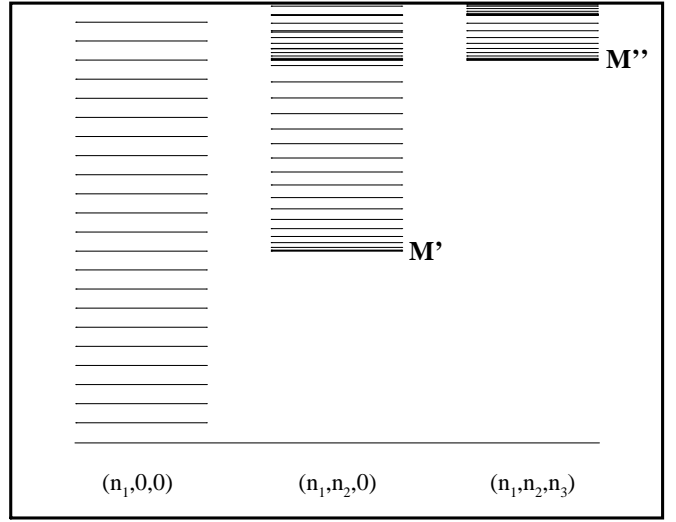}}
\caption{Physical hierarchies in the Kaluza-Klein spectrum.
    For this sketch, we have taken $R_1= 10 R_2 = 20 R_3$ on a rectangular
     torus with $\alpha_{12}= \alpha_{13}= \alpha_{23}=\pi/2$.  The sizes
     of the modular-invariant hierarchies are given by $M'$ and $M''$.} 
\label{KKhierarchy}
\end{figure}

The key to a proper definition of ``hierarchy'' is
to focus purely on the (modular-invariant) Kaluza-Klein spectrum.
Let us consider the case of a three-torus for simplicity;
a sketch with two physical hierarchies is given in 
Fig.~\ref{KKhierarchy}.
If the low-energy Kaluza-Klein spectrum resembles that of a circle,
with a single tower of equally spaced states with masses $k/\rho$, 
and if this pattern exists up to some mass scale $M'$ before additional
unexpected states appear, then we may say that a hierarchy exists
of magnitude $\rho M'\gg 1$.  A similar procedure can be used to
define the hierarchy $M'' \gg M'$ for a third extra dimension, and
so forth.

Note that this also resolves the issue raised above concerning
whether it is legitimate to restrict our attention to states
with vanishing $n_3$ and $n_2$ when passing to lower energy scales.
Anomalously light states with non-vanishing $n_3$ or $n_2$ 
exist only when we have chosen a poor modular ``slice'' 
(analogous to a gauge slice) on which to describe the physics.  
In other words, the lightest states may no longer be those states
for which $n_2$ or $n_3$ vanish.
In all cases, however, the true modular-invariant size of the
hierarchy can always be determined as discussed above, namely 
by taking $M'$ to be the mass scale at which a new tower of states appears.
In a rough sense, 
this is equivalent to defining the hierarchy as the ratio of radii 
on the modular slice for which the sines of all shape angles are maximized
(thereby minimizing the non-diagonal terms in the Kaluza-Klein mass formulas). 
We shall discuss these issues more fully 
in Ref.~\cite{nextpaper}.

A similar remark holds for the 
correspondence relations in Eqs.~(\ref{right1}) and (\ref{right2}).
Strictly speaking, these relations hold as written only 
when the light states are those for which 
$n_3$ and $n_2$ respectively vanish.
As discussed above, this occurs when the sines of the 
shape angles are large {\it or}\/ when the hierarchies in 
the radii are sufficiently large.    
For example, in the case of a two-torus, Eq.~(\ref{right2}) 
holds as written only when $r_1 \sin\theta \gg r_2$. 
On other modular slices, these relations must be modified according 
to Eq.~(\ref{modtrans}).   

Finally, we emphasize that in this paper we are considering 
the compactification geometry as deduced through the Kaluza-Klein 
spectrum.  If string theory is the fundamental theory, then 
winding modes also exist;  whether or not such
modes affect the light spectrum depends
on the relation between the compactification radii 
and the string scale.  If the string scale exceeds
$R_3^{-1}$ where $R_3$ is the smallest radius,
then the winding modes will all be heavy and play no
role in this analysis.  Other configurations
may be more complicated, and will
be discussed in Ref.~\cite{nextpaper}.
Similar remarks also apply for tori with background 
antisymmetric tensor fields~\cite{nextpaper}.

\section{Examples of Shadowing}

Let us now give some examples of the shadowing effects embodied
in Eqs.~(\ref{right1}) and (\ref{right2}).
Our purpose is not to propose a particular set of numerical parameters
for specific phenomenological purposes,
but merely to illustrate the different phenomenological 
possibilities that shadowing provides.  
In each case, we shall assume that the ``true''
compactification manifold is a three-torus and determine how this
three-torus is perceived at various energies.  

Let us begin by assuming 
$R_1^{-1}=10^3$~GeV, 
$R_2^{-1}=10^{11}$~GeV, and $R_3^{-1}=10^{19}$~GeV.
We shall also take $\alpha_{12}=\pi/2$, 
and $\alpha_{13}=\alpha_{23}= \pi/3$.  
The most straightforward way to analyze the Kaluza-Klein
spectrum is to write the Kaluza-Klein masses in Eq.~(\ref{KKthree})
in the ``diagonal'' form
\beq
   \calM^2 ~=~ \sum_{i=1}^3  {\tilde n_i^2\over \tilde R_i^2}~
\eeq
where the eigenvalues are given by
\beqn
       \tilde R_1^{-1}  &\approx&  10^3~{\rm GeV}~,\nonumber\\   
       \tilde R_2^{-1}  &\approx&  10^{11} ~{\rm GeV}~,\nonumber\\   
       \tilde R_3^{-1}  &\approx&  \sqrt{2} \times 10^{19}~{\rm GeV}~ 
\eeqn
and where the eigenvectors are given by 
\beqn
          \tilde n_1 &\approx &   n_1 - (2.5\times 10^{-25})\,n_2 - (5\times 10^{-17})\,n_3 \nonumber\\
          \tilde n_2 &\approx &   (2.5\times 10^{-41})\,n_1 + n_2 - (5\times 10^{-9})\,n_3 \nonumber\\
          \tilde n_3 &\approx &   (5\times 10^{-17})\,n_1 + (5\times 10^{-9})\,n_2 + n_3~.
\eeqn
Given this form, we immediately see that the lowest-lying Kaluza-Klein states 
are those with only $n_1$ non-zero.  For example, if $n_2$ is non-zero, then 
we immediately obtain contributions to $\calM$ in the neighborhood of $10^{11}$~GeV.
Note that it is indeed possible to cancel $\tilde n_2$ 
even if $n_2=1$ by taking $n_1\approx -4\times 10^{40}$ or $n_3 \approx 2\times 10^8$.
However, these values make $\tilde n_1$ or $\tilde n_3$ extremely large,
again inducing contributions to $\calM$ of size exceeding $10^{11}$~GeV.
A similar argument applies for contributions with non-zero $n_3$. 
 From this, we conclude that the Kaluza-Klein spectrum in this example indeed
exhibits two {\it physical}\/ hierarchies:  one between $10^3$~GeV and $10^{11}$~GeV,
and one between $10^{11}$~GeV and $10^{19}$~GeV.
The Kaluza-Klein spectrum in this example
therefore resembles that shown in Fig.~\ref{KKhierarchy}.

The observer with energies below $10^{11}$~GeV
can detect only the $n_1$ excitations.
Adding together the contributions from $\tilde n_1$, $\tilde n_2$, and $\tilde n_3$,
we see that the masses of these states are given by
\beq
        \calM^2 ~\approx~  n_1^2 \, \left( 1.5\times 10^6~{\rm GeV}^2\right)~.
\label{lowmodes}
\eeq
Thus, this observer concludes that the compactification space is a circle
of radius $\rho^{-1}\approx \sqrt{3/2} \times 10^3$~GeV, in accordance with Eq.~(\ref{right2}).
Note that while the above results are only approximate, the result in Eq.~(\ref{right2})
is indeed exact. 
Likewise, the observer with energies below $10^{19}$~GeV sees a two-torus with
radii $r_1^{-1} = (2/\sqrt{3})\times 10^{3}$~GeV,
      $r_2^{-1} = (2/\sqrt{3})\times 10^{11}$~GeV,
and twist angle $\theta\approx 71^\circ$.
This occurs even though the true ``base'' of the original three-torus is completely
rectangular!

In the above example, the numerical distortions of the low-energy parameters relative to the 
parameters of the original three-torus are not large.
However, these distortions are significant, they persist over the whole hierarchy,
and they do not disappear even as the smallest dimension(s) are taken to zero size.
Indeed, the lowest-energy observer sees the regularly spaced 
Kaluza-Klein states in Eq.~(\ref{lowmodes})  stretching over eight
orders in magnitude in energy, yet no corresponding radius of this size 
actually exists in the ``true'' compactification geometry.  It is only the 
presence of two further extra dimensions, many orders of magnitude smaller, 
that causes this distortion!

These shadowing effects become even more dramatic in cases
where we approach a limit $\alpha_{13} + \alpha_{23} \to \alpha_{12}$ 
in the original three-torus.
In such cases, the orientation of the Planck-sized extra dimension associated
with $R_3$ is highly ``squashed'' relative to the two larger dimensions 
associated with $R_1$ and $R_2$.
For example, let us assume $\alpha_{12}=\pi/2$, as before, but 
let us now take $\alpha_{13}=\pi/3 + t$ and $\alpha_{23}=\pi/6  + t$
where $t\ll 1$.
Even though the ``base'' of this three-torus is actually rectangular
with $\alpha_{12}=\pi/2$, 
this base will appear to a low-energy observer as if it 
has a nearly {\it vanishing}\/ shape angle $\theta\sim \sqrt{t}$ 
after the Planck-sized extra dimension is integrated out.
In other words, the squashing of the original Planck-sized 
extra dimension relative to the large dimensions
is perceived by a low-energy observer
as a squashing of the two large 
dimensions with respect to each other!
We stress that this illusion
is wholly due to the existence of the 
third {\it Planck-sized}\/ extra dimension.
Note that even though this example involves ``squashed'' 
extra dimensions, the use of Eqs.~(\ref{right1}) and 
(\ref{right2}) is justified provided the hierarchy of 
radii is sufficiently large compared to the degree of squashing.

Further developing this example, let us 
also assume that the base radii in the original three-torus have 
equal lengths, $R_1 = R_2 $.
We then find effective radii $r_1/r_2 = \sqrt{3}$ for the resulting two-torus.
Remarkably, this result (a nearly squashed two-torus whose
two radii have an algebraic irrational ratio)
are exactly the preconditions needed for the invisibility mechanism 
presented in Ref.~\cite{firstpaper}. 
Indeed, we now see that the large values for the parameter 
$|\tau|$ discussed in Ref.~\cite{firstpaper}
can be realized purely as the result of {\it shadowing}\/ from a three-torus in
which $s_{23}/s_{13}\gg 1$!  

Other interesting phenomenologies are also possible.
For example, even a {\it rectangular}\/ two-torus can sometimes be
nothing but a low-energy illusion;  one configuration that accomplishes this is
to take $\alpha_{12}=\pi/3$, while $\alpha_{13}=\alpha_{23}=\pi/4$.
Thus, even if the perceived shape moduli appear to be trivial to a
low-energy observer, non-trivial shadowing may still be at work in producing
this effect.
 
Non-trivial shape moduli can also be used to 
generate physical hierarchies
when extra dimensions are integrated out.
For example, even if $R_1$ and $R_2$ are of equal magnitude
with $\alpha_{12}=\pi/2$, the third dimension may have
$\alpha_{13}\to 0$.
Even though this third dimension is Planck-sized, its
severe orientation induces a physical hierarchy 
between the two large dimensions.  In other words, a low-energy
observer will observe $r_2/r_1\gg 1$ even
though $R_1=R_2$.  

Further examples and their phenomenological implications
will be discussed in Ref.~\cite{nextpaper}.

\section{Conclusions:  ~Shadowing, \\ Geometry, and Renormalization}

The main result of this paper is that our perception of 
the compactification geometry associated with already-discovered
large extra dimensions can be significantly distorted by the presence of additional,
as-yet-undiscovered smaller dimensions.  As we go to higher and
higher energies and discover these additional dimensions, 
our description of the compactification manifold changes --- 
not merely in its dimensionality but also in the radii and shape
angles that parametrize {\it all length scales}\/ of this geometry.
Indeed, our perception of very simple geometric quantities 
such as the radii and shape moduli associated with the largest 
(and experimentally accessible) extra dimensions continually evolves 
as a function of the energy 
with which we probe this manifold --- even though the largest
extra dimensions are already detected and their geometric properties 
are already presumed known.

Of course, this is not a new concept in physics:  this is nothing
but {\it renormalization}\/.  Thus, in this sense, we see
that the apparent compactification geometry is not fixed at all, but
rather undergoes renormalization much like other ``constants'' of nature.
Indeed, the correspondence relations in Eqs.~(\ref{right1}) 
and (\ref{right2}) serve as ``renormalization-group equations''
which describe the flow of the perceived geometric parameters
associated with the largest extra dimensions
as we pass through the thresholds associated with additional, smaller
extra dimensions.
Moreover, as we discussed, this renormalization-group evolution 
cannot be undone through modular transformations.  This evolution
is therefore a truly physical effect, 
one which corresponds to perceived changes in {\it topology}\/ 
as well as geometry. 

It is important to understand the precise sense in which 
shadowing can be considered as renormalization.
Clearly, in different energy ranges, we are employing
a series of different effective field theories in order to describe
the Kaluza-Klein spectrum:  the effective field theory
at the lowest energy scales has a single parameter
$\rho$; the effective field theory at intermediate energies
has three parameters $(r_1,r_2,\theta)$;  and the effective
field theory at still higher energies has six 
parameters $\lbrace R_i,\alpha_{ij}\rbrace$. 
The correspondence relations in Eqs.~(\ref{right1}) and
(\ref{right2}) are thus properly viewed as matching conditions 
(or threshold relations) between different
effective field theory descriptions of the same physics
at different energy scales.
Indeed, these matching conditions reflect nothing more than our
requirement that the physical Kaluza-Klein spectrum remain 
invariant as we change our description of the physics 
by changing the cutoffs inherent in our 
sequence of effective field theories.

However, the cumulative effect of such threshold corrections 
as we pass between different effective theories is 
precisely what is usually meant by 
renormalization.  Indeed, if we imagine extrapolating our calculations 
to incorporate a continuing series of hierarchies  
corresponding to a continuing series of extra dimensions,
then the corresponding series of matching conditions constitutes a 
renormalization group ``flow''.  Under this flow, the values of 
parameters such as the radius of the largest extra dimension, 
be it $\rho$ or $r_1$ or $R_1$, evolve in a non-trivial way due 
to the presence of non-trivial shape moduli.  
In other words, such parameters are renormalized.
     
We stress that this ``renormalization'' is a purely classical
effect, one which arises for purely geometric reasons.
As such, it does not incorporate further quantum-mechanical effects
which may arise due to the quantum field-theoretic renormalization
of the Kaluza-Klein masses.
Indeed, implicit in our previous discussions has been the assumption 
that the Kaluza-Klein mass spectrum is itself a 
physical observable, one which can be measured independently of other
parameters in the theory.  
Of course, in a more general interacting theory,
the renormalization flow of the 
compactification geometry may receive further quantum-mechanical contributions.

We also stress that in this paper we have considered only the
simplest case of flat, toroidal compactifications without torsion.
In principle, one can also study more complicated manifolds
with more complicated Kaluza-Klein spectra~\cite{Kaloper}.

In all cases, however, our main observation stands:  
quantities such as compactification radii --- quantities which
one might have na\"\i vely assumed to be fixed once the
corresponding extra dimensions are discovered --- are not
fixed at all.  Instead, they are effectively renormalized as we
pass to higher and higher energies and as additional extra 
dimensions become apparent.  
Since one can never be satisfied experimentally that one has
discovered the totality of possible extra dimensions, this
process need not terminate.  It then becomes experimentally
meaningless to speak of a ``true'' compactification geometry,
in exactly the same way as it is meaningless to speak
of the ``true'' electron charge. 

There have been several recent discussions relating compactification
geometry and renormalization.
These include the ``deconstruction'' idea~\cite{decon}, as well as the
AdS/CFT correspondence~\cite{Maldacena}
in the context of higher-dimensional models
with localized gravity~\cite{RS}. 
Yet each of these cases is quite different from the shadowing 
effect we are discussing here.
In the localized gravity/AdS case, the apparent compactification geometry is fixed;
what changes is the renormalization scale on a particular brane as it
moves through the AdS geometry.
Likewise, in the deconstruction case, extra dimensions are
generated as the result of certain fields condensing; 
this change has nothing to do with the geometry of the space itself.
By contrast, our results hinge purely on the geometric
properties of the compactification space and its
manifestations at different energy scales.

The implications of the shadowing effect are likely to be profound.
Rather than think of compactification geometry as fixed and immutable,
we instead must think of it as something renormalizable.
This clearly raises a number of provocative questions.
What are the properties of the renormalization flow of compactification 
geometry as we approach the fundamental scale of quantum gravity
where the whole notion of a continuous spacetime might break down?
If string theory is the correct underlying theory,
how can we incorporate winding modes (and ultimately T-duality) into this picture?
Conversely, might our own four-dimensional spacetime only {\it appear}\/ to be
large and flat as a consequence of shadowing from additional spacetime 
dimensions?  Might this provide a new approach to the cosmological constant
problem?
Indeed, in what sense is spacetime geometry knowable at all --- are
there analogues of renormalization-group invariants?

There is an old question in mathematical physics:  Can one 
hear the shape of a drum?  Clearly, our answer is that 
drums have no absolute shape.  Instead, the shape of the drum 
depends on how well one listens.

\section*{Acknowledgments}

This work was supported in part by the National Science Foundation
under Grant PHY-0071054, and by a Research Innovation Award from 
Research Corporation. 
We wish to thank U.~van~Kolck for valuable discussions.
 

\medskip



\begin{references}

\bibitem{firstpaper}   K.R.~Dienes, hep-ph/0108115 
             ({\it Phys.\ Rev.\ Lett.}\/, in press).
\bibitem{nextpaper}   K.R.~Dienes and A.~Mafi, 
         {\it Compactification on Manifolds with Non-Trivial Shape Moduli}\/, to appear.
\bibitem{Kaloper}
        See, {\it e.g.}\/, N.~Kaloper {\it et al.},
          Phys.\ Rev.\ Lett.\  {\bf 85} (2000) 928
        [hep-ph/0002001].
\bibitem{decon} N.~Arkani-Hamed, A.G.~Cohen and H.~Georgi,
              Phys.\ Rev.\ Lett.\  {\bf 86} (2001) 4757
                  [hep-th/0104005];
            C.T.~Hill, S.~Pokorski and J.~Wang, hep-th/0104035.
\bibitem{Maldacena}
          J.~Maldacena,
           Adv.\ Theor.\ Math.\ Phys.\  {\bf 2} (1998) 231
             [hep-th/9711200].
\bibitem{RS}
        L.~Randall and R.~Sundrum,
         Phys.\ Rev.\ Lett.\  {\bf 83} (1999) 3370
          [hep-ph/9905221];
         Phys.\ Rev.\ Lett.\  {\bf 83} (1999) 4690
         [hep-th/9906064].


\end{references}
\end{document}